\documentclass[aps,pre,showpacs,twocolumn,groupedaddress,superscriptaddress]{revtex4}
\usepackage{graphicx}%
\usepackage{amsmath,citesupernumber}
\usepackage{amssymb}
\usepackage{dcolumn}
\usepackage{epsfig}
\usepackage{graphics}
\usepackage{bm}

\begin{document}

\title{Adaptive coupling induced multi-stable states in complex networks}

\author{V.~K.~Chandrasekar}%
 \email{chandru25nld@gmail.com}
\affiliation{Centre for Nonlinear Dynamics, School of Physics,
Bharathidasan University, Tiruchirappalli - 620 024, Tamilnadu, India}

\author{Jane H.~Sheeba}%
\email{jane.sheeba@gmail.com}
\affiliation{Centre for Nonlinear Dynamics, School of Physics,
Bharathidasan University, Tiruchirappalli - 620 024, Tamilnadu, India}

\author{B.~Subash}%
\email{subash.udt@gmail.com}
\affiliation{Centre for Nonlinear Dynamics, School of Physics,
Bharathidasan University, Tiruchirappalli - 620 024, Tamilnadu, India}

\author{M.~Lakshmanan}%
\email{lakshman@cnld.bdu.ac.in}
\affiliation{Centre for Nonlinear Dynamics, School of Physics,
Bharathidasan University, Tiruchirappalli - 620 024, Tamilnadu, India}

\author{J.~Kurths}%
\email{kurths@pik-potsdam.de}
\affiliation{Potsdam Institute for Climate Impact Research, 14473 Potsdam, Germany}

\date{\today}

\begin{abstract}
Adaptive coupling, where the coupling is dynamical and depends on the behaviour of the oscillators in a complex system, is one of the most crucial factors to control the dynamics and streamline various processes in complex networks. In this paper, we have demonstrated the occurrence of multi-stable states in a system of  identical phase oscillators that are dynamically coupled. We find that the multi-stable state is comprised of a two cluster synchronization state where the clusters are in anti-phase relationship with each other and a desynchronization state. We also find that the phase relationship between the oscillators is asymptotically stable irrespective of whether there is synchronization or desynchronization in the system. The time scale of the coupling affects the size of the clusters in the two cluster state. We also investigate the effect of both the coupling asymmetry and plasticity asymmetry on the multi-stable states. In the absence of coupling asymmetry, increasing the plasticity asymmetry causes the system to go from a two clustered state to a desynchronization state and then to a two clustered state. Further, the coupling asymmetry, if present, also affects this transition. We also analytically find the occurrence of the above mentioned multi-stable-desynchronization-multi-stable state transition.  A brief discussion on the phase evolution of nonidentical oscillators is also provided. Our analytical results are in good agreement with our numerical observations.
\end{abstract}



\maketitle

\section{Introduction}

Synchronization is one of the most commonly occurring phenomenon in various physical and biological networks and it has been a topic of active research in recent years~\cite{Winfree:67,Kuramoto:84,Pikovsky:01, Strogatz:01}. For instance, synchronization is one of the most crucial dynamical aspects in social networks, neuronal networks, cardiac pacemakers, circadian rhythms, ecological systems, power grids, etc \cite{Strogatz:01, Singer:99, Jane:08b, Timmermann:03, Percha:05, Pfurtschelle:99, Jane:09}. The emergence of collective oscillations in these kinds of coupled systems can be described by the Kuramoto model, which is a mathematically tractable model \cite{Kuramoto:84,Acebron:05}. In this model, when the oscillators are uncoupled they oscillate at their own frequencies. When the oscillators are coupled by the sine of their phase differences, the frequencies of the oscillators are modified leading to synchronization depending upon the coupling strength.

Most of the previous studies in this area \cite{Albert:02, Boccaletti:06, Wu:10, Perc:07} have been devoted to the case where the coupling strength between the different oscillators in the network is fixed. However, considering the self-organizational nature of most complex systems (like that of social systems and neuronal networks), inclusion of an adaptive coupling among the oscillators appears to be more realistic \cite{Seliger:02,Zanette:04,Zimmermann:04,Pacheco:06,Masuda:07,Maistrenko:07,Cateau:08,Chen:08,Gilson:09,Szolnoki:09a,Szolnoki:09b,Poncela:09,Perc:10,Iwasa:10a,Iwasa:10b}.

The emergence of synchronization in complex networks, for example social networks involving opinion formation or brain, is due to the fact that coupling between different entities in the network varies dynamically \cite{ Bi:98,Markram:97,Caporale:08}. The coupling usually varies due to the interplay between the dynamical states and the network topology \cite{Ren:07,Hou:10}. Such networks are called adaptive networks. The presence of adaptive coupling is one of the most crucial factors that control the dynamics and streamline various processess in complex networks. For instance in the brain, where synchronization is not always desirable, the presence of adaptive coupling can lead to desynchronization by the alteration of coupling naturally when the system senses the occurrence of such undesirable synchronization \cite{Chandrasekar:10}. Very recently Aoki and Aoyagi have found that systems with adaptive coupling give rise to three basic synchronization states due to the self-organizational nature of the dynamics \cite{Aoki:09,Aoki:11}. They have discovered the existence of a two-cluster state, a coherent state (in which the oscillators are having a fixed phase relationship with each other), and  a chaotic state in which the coupling weights and the relative phases between the oscillators are chaotically shuffled. This implies that the existence of adaptive coupling accounts for the feature rich, yet self-organized behaviour of real-world complex networks.

What is more interesting in the context of social and biological networks is their ability to adapt themselves and learn from their own dynamics, given that there is no one to administrate their activity and control them \cite{Yuan:11, Zhu:10}. For example, opinion formation in social networks is a complex phenomenon which not only depends on the number of conformists and contrarians \cite{Lama:05} but also on how the opinion evolves in time. This opinion evolution is essentially the cause of self-organization which gives rise to social clusters/groups and also synchronization in the network \cite{Guven:10,Pluc:06}. This kind of adaptive evolution of opinion not only plays a crucial role in the social network but is also found to influence the dynamics of autocatalytic interactions (where the chemical reaction rates are dynamic and depend on the reaction products), biological networks \cite{Furusawa:03,Jain:01,Tero:10,Harris:03,Gross:08}, and so on. 

In this paper, motivated by the above mentioned facts, we explore the role played by adaptive coupling on the synchronization in coupled oscillator systems. We find that adaptive coupling can induce the occurrence of multi-stable states in a system of coupled oscillators. We also find that the weight of the coupling strength and the plasticity of the coupling play a crucial role in controlling the occurrence of multi-stable states. Further, we also find that the effect of asymmetry on the multi-stable states is such that it can drive the system from a multi-stable via phase desynchronized to a multi-stable state. Here by a desynchronized state we essentially mean a coherent state as identified by Aoki and Aoyagi \cite{Aoki:09,Aoki:11} which corresponds to a fixed phase relationship between different oscillators as a function of time. However, in this state the oscillators are also distributed over the entire range $(0,2\pi)$ and so the corresponding state is also called as a phase desynchronized state in the literature \cite{Stro:08,Hong:11,Hong:11e}. In the following we use this later terminology with the understanding that it can also be described as coherent state.

The plan of the paper is as follows: In the following Sec. \ref{model} we introduce the model of identical coupled phase oscillators which we take into consideration. We also give details about the numerical simulation we use. In Sec. \ref{multi} we demonstrate the occurrence of multi-stable states in the model we consider. We first (Sec. \ref{Amulti}) show the absence of multi-stable states in the system without coupling plasticity. We then introduce coupling plasticity (Sec. \ref{Bmulti}) and show how two different synchronization states coexist in the system. In Sec. \ref{Cmulti} we demonstrate how  the coupling time scale influences the occurrence of multi-stable states. In Sec. \ref{phase lag} we introduce a phase asymmetry in the system and demonstrate how it takes the system from a multi-stable state through a desynchronized state to a multi-stable state. In Sec. \ref{ana}, we provide an analytical basis for our numerical results. In Sec. \ref{nio} a brief discussion on the phase evolution in the case of nonidentical oscillators is given. Finally we present the summary of our findings in Sec. \ref{conc}.

\begin{figure*} 
\begin{center} 
\includegraphics[width=15.50cm,height=17.50cm]{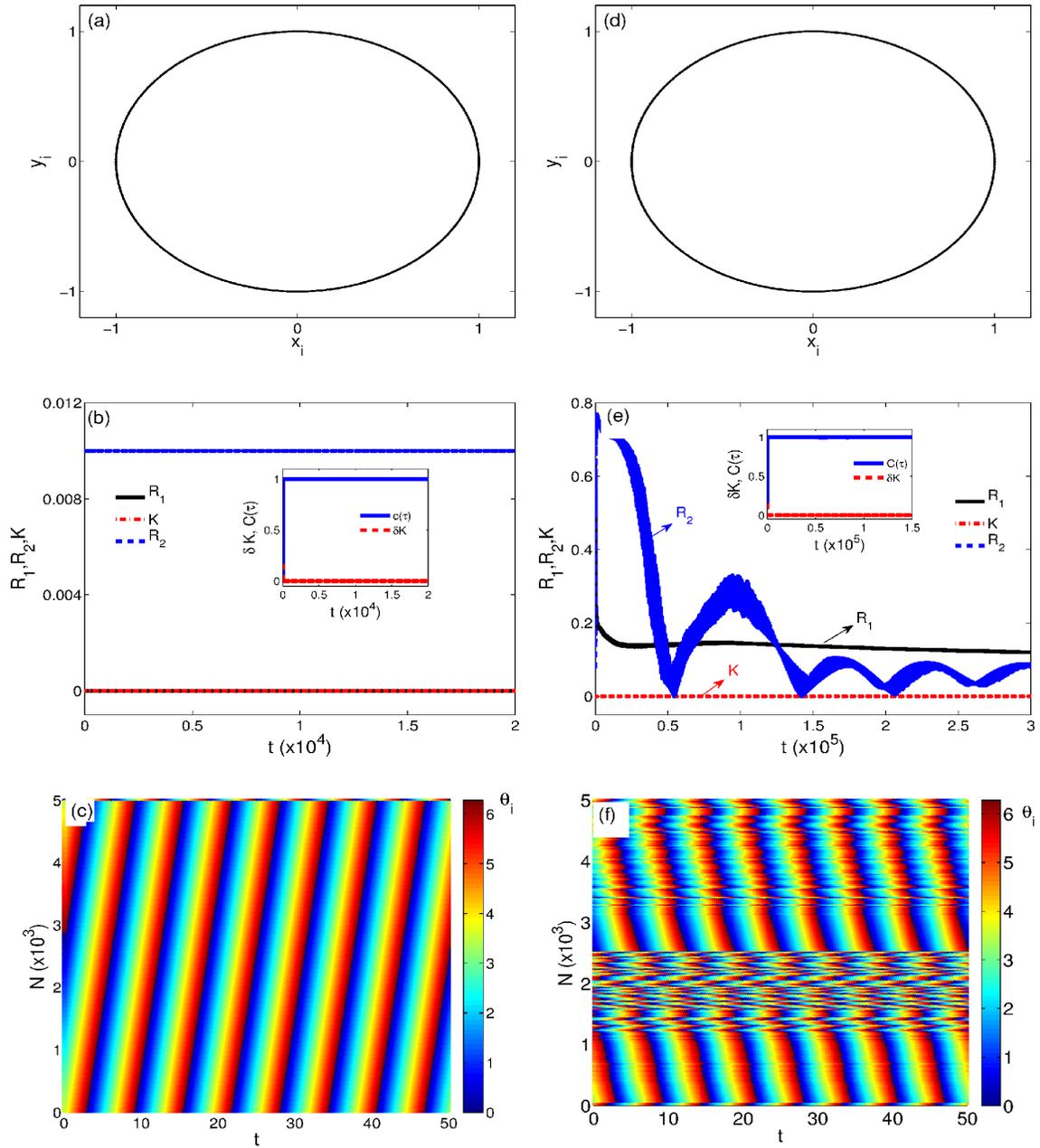} \caption{In the absence of adaptive coupling ($\eta=0.0$), system (3) exhibits similar behaviour of phase desynchronization for two different initial conditions. Panels (a)-(c) in the left column are plotted for a uniform distribution of initial phases  (i.e. phases are distributed uniformly between 0 to 2$\pi$  for N=5000 oscillators, see the text for more details) and panels (d)-(f)  in the right column are plotted for a random distribution of initial phases (i.e. phases are distributed randomly between 0 and $\pi$). Panels (a) and (d) are the phase space ($((x_i,y_i)=(\cos\theta_i,\sin \theta_i))$) portraits. In panels (b) and (e), the time evolution of the order parameters $R_1$ (solid black line), $R_2$ (dashed blue line) and the average of coupling weights $K$ (dot-dashed red line) are plotted. The insets of panels (b) and (e) correspond to the rate of change of $K$, namely $\delta K$ (dashed red line), and the autocorrelation $C(\tau)$ (solid blue line). In panels (c) and (f) the asymptotic time evolution of the oscillator phases are plotted  along with their position (N) in the network (The colour coding $(0,2\pi)$ elucidates the phases of each oscillator at every time step). The parameters in (3) are chosen as $\omega =1.0$, $\sigma=1.0$ and $N=5000$. In panels (a), (c), (d) and (f) all the data are plotted after sufficient transients (of the order $10^6$) are left out (see also the text).}\label{fig1}
\end{center}
\end{figure*}

\section{The model}
\label{model}
Let us consider a system of coupled phase oscillators described by the following Kuramoto-type~\cite{Kuramoto:84} evolution equations
\begin{eqnarray}
\label{cho01}
\dot{\theta_i}= \omega- \frac{K}{N}\sum_{j=1}^{N}
f(\theta_i-\theta_j),\; i=1,2, \ldots, N,
\end{eqnarray}
where $\theta_i$ is the phase of the $i$th oscillator and $f$ is a $2\pi$-periodic coupling function.  Here $\omega$ is the natural frequency of the oscillators and $K$ is the coupling strength between the oscillators. If $K$ is positive, it denotes attractive interaction between the oscillators and if $K$ is negative, it implies repulsive interaction. Assuming that a given population of coupled oscillators in real life systems have both kinds of couplings (attractive and repulsive or equivalently contrarian and conformist), we replace $K$ by $\sigma k_i$, where $\sigma$ represents the strength of the coupling between the oscillators in the system and the $k_i$ are the coupling weights between the oscillators. $K$ can also be assumed to have spin glass-type coupling $k_{ij}$. However in this type of coupling there will be no room for characterizing each oscillator in the system by an intrinsic property \cite{Hong:11}. Nevertheless, models with such spin glass type interactions have been extensively studied by various authors \cite{Aoki:11, Seliger:02, Ren:07, Hou:10, Yuan:11, Wu:11, Huang:06, Li:11, Zhu:10}.

We consider the coupling weights $k_i$ to be dynamic, represented by the following dynamical equation
\begin{eqnarray}
\label{cho02}
\dot{k_i}= \frac{\eta}{N}\sum_{j=1}^{N}
g(\theta_i-\theta_j),\; i=1,2, \ldots, N,
\end{eqnarray}
where $g$ is a $2\pi$-periodic function and can be called as the plasticity function \cite{Aoki:11} which determines how the coupling weight depends on the relative timing of the oscillators. $\eta$ is the plasticity parameter. The evolution of the coupling is slower than the evolution of the oscillators when $\eta<<1$ and the time scale of the coupling dynamics is given by $\eta^{-1}$. Here we choose the simplest possible periodic functions for $f$ and $g$ as $f(\theta)=\sin(\theta)$ and $g(\theta)=\cos(\theta)$. Such a coupling implies the fastest learning configuration in the system, that is, when the oscillators are in phase, the coupling coefficient grows fastest and when the oscillators are out-of-phase, the coupling coefficient decays fastest (Hebbian-like).
With this choice, the model equations can be rewritten as
 \begin{eqnarray}
\label{cho03}
\dot{\theta_i}&=& \omega - \frac{\sigma}{N}\sum_{j=1}^{N}
k_i\sin(\theta_i-\theta_j),\nonumber\\
\dot{k_i}&=& \frac{\eta}{N}\sum_{j=1}^{N}
\cos(\theta_i-\theta_j),\; i=1,2, \ldots, N.
\end{eqnarray}
Synchronization in the system can be quantified using the Kuramoto order parameter,
\begin{eqnarray}
\label{cho01a}
Z_1=R_1e^{i\psi_1}= \frac{1}{N}\sum_{j=1}^Ne^{i\theta_j}=\frac{1}{N}\sum_{j=1}^N(x_i+iy_i).
\end{eqnarray}
\begin{figure*}
\begin{center}
\includegraphics[width=15.50cm,height=17.5cm]{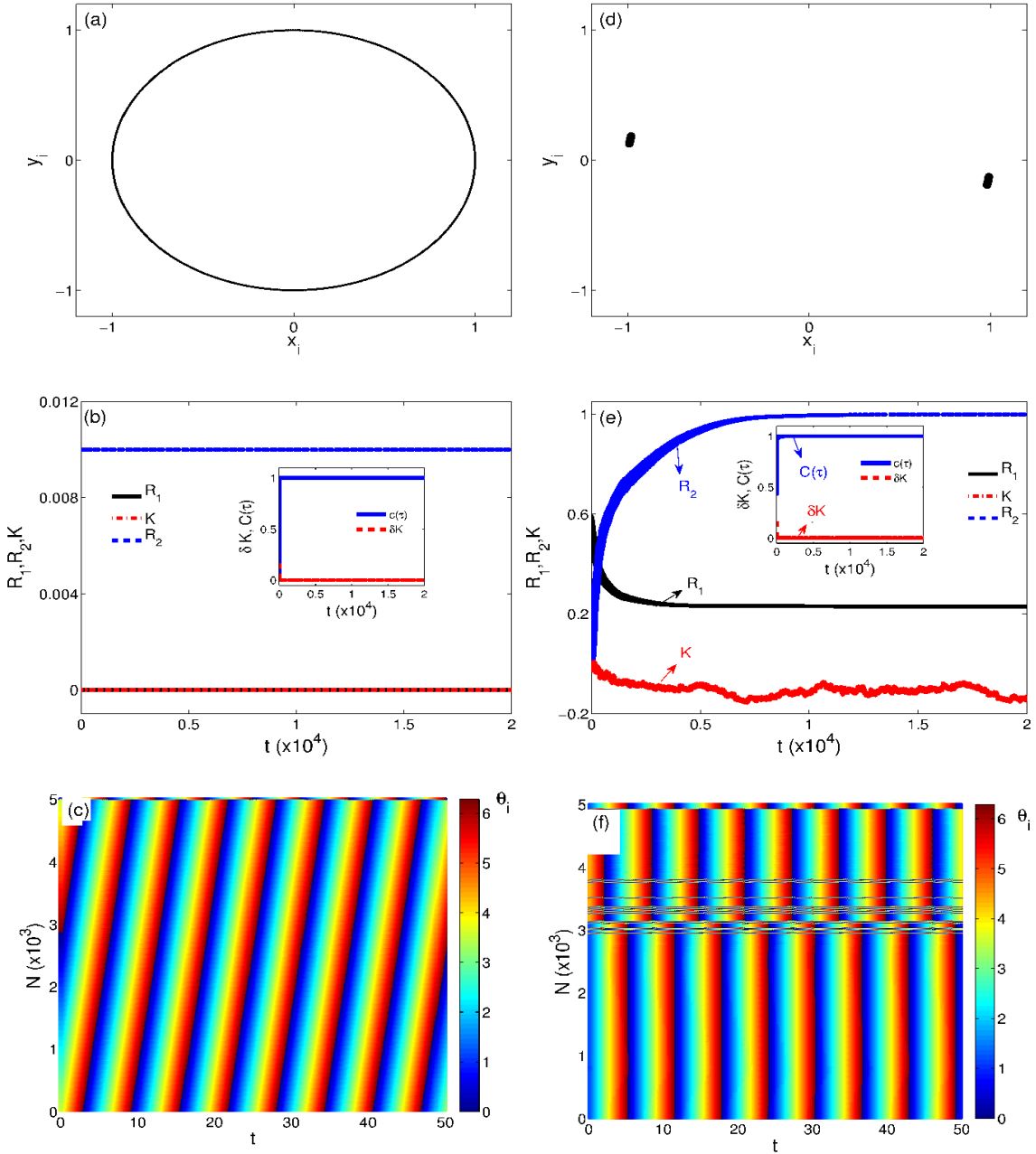}
\caption{All the panels and insets correspond to those in Fig. \ref{fig1} but now in the presence of adaptive coupling ($\eta=0.05$). The system exhibits different synchronization dynamics for different initial conditions. For uniformly distributed initial phases the system exhibits phase desynchronization (panels (a)-(c)) and for a random distribution of initial phases there exist a two-cluster state in the system (panels ((d)-(f)). In panel (f) the oscillator phases are plotted for a longer period of time than that of (c) to clearly illustrate the presence of the two cluster state.}
\label{fig2}
\end{center}
\end{figure*}


Here $R_1$ is the coherence parameter that represents the strength of synchronization in the system. A complete synchronization in the system corresponds to $R_1=1$, while complete phase desynchronization (that is, every oscillator in the system has a corresponding oscillator that is in anti-phase synchronization with it) corresponds to $R_1=0$. When there is a partial synchronization in the system, $R_1$ takes a value between $0$ and $1$ and quantifies the strength of synchronization; that is, the value of $R_1$ is directly proportional to the number of oscillators that are in synchrony. 

In our simulations, we consider $N=5000$ oscillators, (except in Fig. \ref{fig7} where we take N=1000). We use a fourth order Runge Kutta routine to numerically simulate the system and we fix the time step to be 0.01.  We have discarded the first $10^6$ time steps and continued our simulations for another $10^3$ time steps, though in some figures (Figs. \ref{fig1}(b,e), \ref{fig2}(b,e), \ref{figp}, \ref{fig4}(a,b), \ref{fig6}(a,b,c),  \ref{fig11}, \ref{fignon}  and \ref{fig8}) we also indicate the effect of transients. The results are shown in the various figures in the text for a small window of time whenever transients are not shown (which we label to start with 0 in the figures for convenience) towards the end of the total simulation time. 
In all our numerical simulations, the initial values of coupling weights $k_i$ are uniformly distributed in [-1,1] by imposing the condition $|k_i|\leq1$ so that whenever the value of $k_i$ goes outside the interval [-1,1] as it evolves, it is immediately brought back to the bound value in the interval.

In the following Section let us explore the changes in the synchronization dynamics of system (\ref{cho03}) as affected by the plasticity in the coupling.

\section{The effect of coupling plasticity}
\label{multi}
We find that multi-stable synchronization states are generated by the system due to the presence of coupling plasticity. In order to demonstrate the same, first let us consider system (\ref{cho03})  without coupling plasticity.

\subsection{Absence of coupling plasticity}
\label{Amulti}
When there is no coupling plasticity in the system, Eq. (\ref{cho02})  becomes 
 \begin{eqnarray}
\label{cho03a}
\dot{k_i}&=& 0  \;\;\mbox{and} \;\;k_i=k_i(0),\; i=1,2, \ldots, N,
\end{eqnarray}
where $k_i(0)$ are the integration constants which are nothing but the initial conditions and are uniformly distributed in [-1,1]. With this coupling function, let us consider the dynamics of system (\ref{cho03}), to begin with.

In Fig. \ref{fig1}, the left column ((a)-(c)) and the right column ((d)-(f)) panels are plotted for two different initial conditions, namely uniform and random distributions for the initial oscillator phases, respectively. For the uniform distribution of initial phases, we have set the values as uniformly distributed  between $0$ and $2\pi$ among all the N oscillators. As an example, we choose the initial phase of the N=1 oscillator as $0$ and then for the subsequent oscillators the phase is increased in units of $2\pi/(N-1)$ so that for the $N^{th}$ oscillator it is $2\pi$.  (We have also checked that the results are invariant for another set of initial conditions corresponding to a uniform distribution of intial phases between $-\pi$ and $+\pi$). On the other hand for the random state of initial phases we have used a random number generator which generates random numbers with normal distribution between 0 to $\pi$. (Again we have confirmed similar results for another set of random initial conditions in the range [-1,+1]). We have plotted in Fig. \ref{fig1} the phase space $((x_i,y_i)=(\cos\theta_i,\sin \theta_i))$ evolution of the oscillators in panels (a) and (d) which shows every oscillator has a corresponding oscillator that is in anti-phase relationship with it, and hence this system is in a phase desynchronized state, the time evolution of the order parameters $R_1$, $R_2$, and the average of coupling weights $K$ (defined below) in panels (b) and (e), and the time evolution of the oscillator phases in panels (c) and (f). Here $R_1=|\frac{1}{N}\sum_{j=1}^Ne^{i\theta_j}|$ is the order parameter of the whole system, while $R_2$ is the order parameter of the two cluster state (where the oscillators are grouped into two clusters) given as $R_2=|\frac{1}{N}\sum_{j=1}^Ne^{2i\theta_j}|$. When $R_2\sim 1$, the oscillators converge to a state of two synchronized clusters that are in antiphase relationship with each other. That is there are two groups of oscillators with phases $\theta$ and $\theta\pm\pi$. In this state, the order parameter $R_1$ is given as
\begin{eqnarray}
R_1=\frac{|N_1(\theta)-N_2(\theta\pm\pi)|}{N}=\Delta N,
\end{eqnarray}
where $N_1+N_2=N$ and $N_1(\theta)$ and $N_2(\theta\pm\pi)$ are the number of oscillators phase locked in $\theta$ and $\theta\pm\pi$, respectively. If $R_2\sim 1$ when $R_1\sim 0$ this means that the two clusters have equal number of oscillators in them ($N_1=N_2$). On the other hand if $R_2\sim 1$ when $R_1>0$ then the clusters contain different number of oscillators. In panels (b) and (e) the solid, dashed and dot-dashed lines represent $R_1$, $R_2$ and $K$, respectively. Here $K$ is the average of the coupling weights $k_i$ given as
\begin{eqnarray}
\label{cop}
K= \frac{1}{N}\sum_{j=1}^N k_j.
\end{eqnarray}
We have in addition plotted the average rate of change of the coupling weights $\delta K$ (over all the connections), 
\begin{eqnarray}
\label{cop1}
\delta K= \frac{1}{N}\sum_{j=1}^N \frac{|k_j(t)-k_j(t-\delta)|}{\delta},
\end{eqnarray}
and the auto-correlation function of the oscillator phases,
\begin{eqnarray}
\label{ac1}
C(\tau)= <|\frac{1}{N}\sum_{j=1}^N e^{i(\theta_j(t)-\theta_j(t-\tau))}|>,
\end{eqnarray}
as insets in panels (b) and (e).

In Figure \ref{fig1}, it is obvious that the oscillators remain in the desynchronized state irrespective of the initial conditions, in the absence of coupling plasticity. The phase portraits in panels (a) and (d) show that the oscillators are uniformly distributed on a unit circle, implying desynchronization. This is also confirmed by the time evolution of the phases shown in panels (c) and (f). The order parameters $R_1$ and $R_2$ also take values close to 0 in this case. $K$ and $\delta K$ remain zero, since all the oscillators are uniformly distributed in [-1,1] and the correlation remains at 1 indicating that this desynchronized state is a steady state; that is the phase relation between any two oscillators in the system is fixed for all times.
The transient behaviour of the order parameters $R_1$, $R_2$ and the coupling strength $K$ for the case of no coupling plasticity is also included in panels (b) and (e). In the absence of coupling plasticity the system takes much longer period of time to reach the asymptotic regime in the case of random distribution of initial phases.

\subsection{The presence of coupling plasticity (Eq. (\ref{cho03}))}
\label{Bmulti}
\begin{figure}
\begin{center}
\includegraphics[width=8.0cm,height=7.0cm]{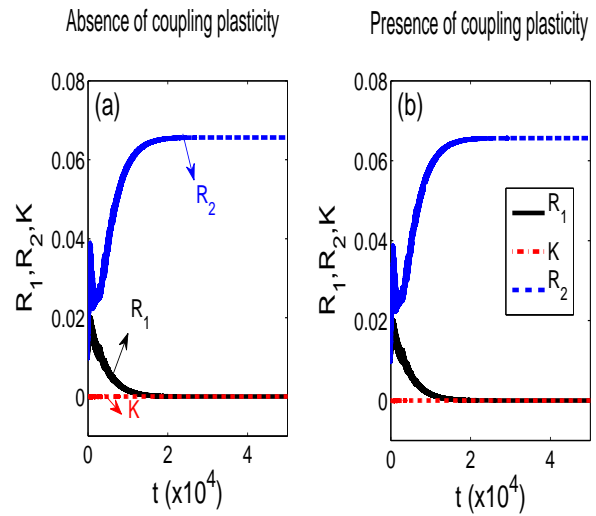}
\caption{ Desychronized states ($R_1$=0, $R_2\sim0$) of the oscillators both in the case of (a) absence ($\eta=0.0$) and (b) presence ($\eta=0.05$) of coupling plasticity for a slightly perturbed set of initial phases distributed uniformly between $0.1\pi$ to $1.9\pi$ exhibiting similar behaviour of desynchroization as in panels 
(a), (b), (c) in Figs. \ref{fig1} and \ref{fig2}. }
\label{figp}
\end{center}
\end{figure}

Now let us introduce plasticity in the adaptive coupling in the system. In this case, the dynamics of the coupling weights depends upon the phase relation between the oscillators, and hence the initial phases of the oscillators strongly affect the synchronization states in the system. We have plotted Fig. \ref{fig2} exactly in the same manner as Fig. \ref{fig1} except that we have now introduced the coupling plasticity in the system with the strength $\eta=0.05$. Here we see that for a uniformly distributed initial oscillator phases, the left panels (a,b,c) resemble that of Fig. \ref{fig1}, that is, the oscillators remain phase desynchronized. However in the right panels (d,e,f), when the initial phases are distributed randomly, we clearly see the emergence of a two-cluster synchronization state where the clusters are in antiphase relationship with each other, which is confirmed by $R_2\sim1$. Since $R_1\neq0$, even though the two clusters are in antiphase relationship, this implies that $\Delta N\neq0$. In short, the two clusters do not have an equal number of oscillators in them. However the autocorrelation remains constant at a value close to 1 for both the initial conditions indicating that the phase relationship between the oscillators is asymptotically stable irrespective of whether the oscillators are synchronized or desynchronized. We call the existence of a desynchronized state and a two-cluster state as a multi-stable regime \cite{Aoki:11a}. In Figure \ref{fig2}(e), we have also shown the transient behaviour of order parameters $R_1$, $R_2$ and the coupling strength $K$ in the presence of adaptive coupling. By comparing panel (e) of Figures \ref{fig1} and \ref{fig2}, it is clear that the presence of adaptive coupling enhances the system to reach the stable states in a much shorter period of time in the case of random initial conditions. Also in Fig. \ref{figp}, we have plotted the time evolution of the order parameters $R_1$ and $R_2$ as well as the average coupling $K$ (a) in the absence ($\eta=0.0$) and (b) in the presence ($\eta=0.05$) of dynamic coupling $\eta$ for a slightly perturbed set of uniform initial conditions in the range $0.1\pi$ to $1.9\pi$ (instead of $0$ to $2\pi$ discussed earlier). We find that the system continues to be in a phase desynchronized state asymptotically, showing the stable nature of the underlying dynamics.

\begin{figure*}
\begin{center}
\includegraphics[width=16.5cm,height=9.0cm]{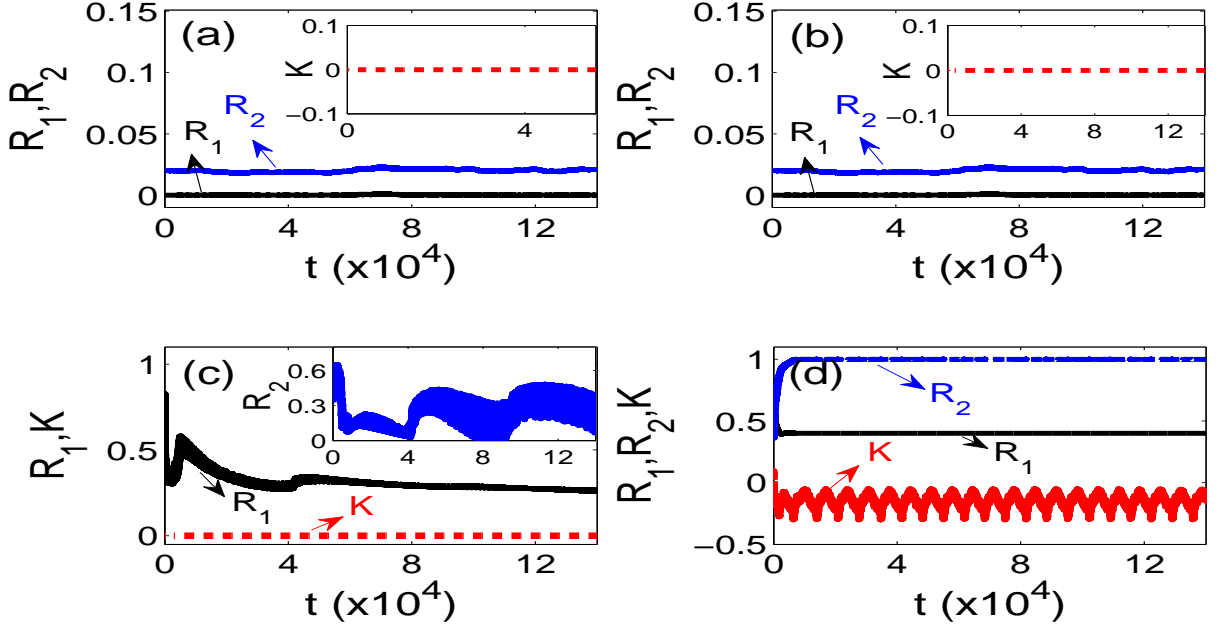}
\caption{ Under the influence of independent external noise $\zeta_i$ as given in Eq. (\ref{cho_nois}), the system exhibits similar behaviour of multistability state as in the case of absence of external perturbation (Figs. \ref{fig1} and \ref{fig2}). The order parameters $R_1$, $R_2$ and the coupling strength $K$ are plotted for both uniform (panels (a), (b)) as well as for random distribution (panels (c), (d)) of initial phases in the cases of (1) absence of adaptive coupling ($\eta=0.0$): panels (a), (c),  and (2) presence of adaptive coupling ($\eta=0.05$): panels  (b), (d). Here the noise strength $D=0.0001$.}
\label{fign}
\end{center}
\end{figure*}

Further, we have also included a gaussian white noise in the model Eq. (\ref{cho03}) to study the stable nature of the multistable state in the presence of external perturbations. In this case the model Eq. (\ref{cho03}) becomes
\begin{eqnarray}
\label{cho_nois}
\dot{\theta_i}&=& \omega - \frac{\sigma}{N}\sum_{j=1}^{N} 
k_i\sin(\theta_i-\theta_j)+ \zeta_i(t),\nonumber\\
\dot{k_i}&=& \frac{\eta}{N}\sum_{j=1}^{N}
\cos(\theta_i-\theta_j),\; i=1,2, \ldots, N,
\end{eqnarray} 
where the variables $\zeta_i$ correspond to independent white noise that satisfies $<\zeta_i(t)>$=0, and  
$<\zeta_i(s)\zeta_j(t)>=2D\delta_{ij}\delta(s-t)$. Here D is the noise strength. For our study we choose D=0.0001. In Fig. \ref{fign}, we have briefly presented the order parameters $R_1$, $R_2$ and the average coupling strength $K$ as a function of time in the presence of the above external noise for both the cases of absence and presence of adaptive coupling. On comparing the present dynamics with Figs. \ref{fig1} and \ref{fig2} corresponding to the absence of noise, we can easily see that the qualitative nature of the dynamics remains unchanged.

\begin{figure}
\includegraphics[width=8.5cm,height=6.0cm]{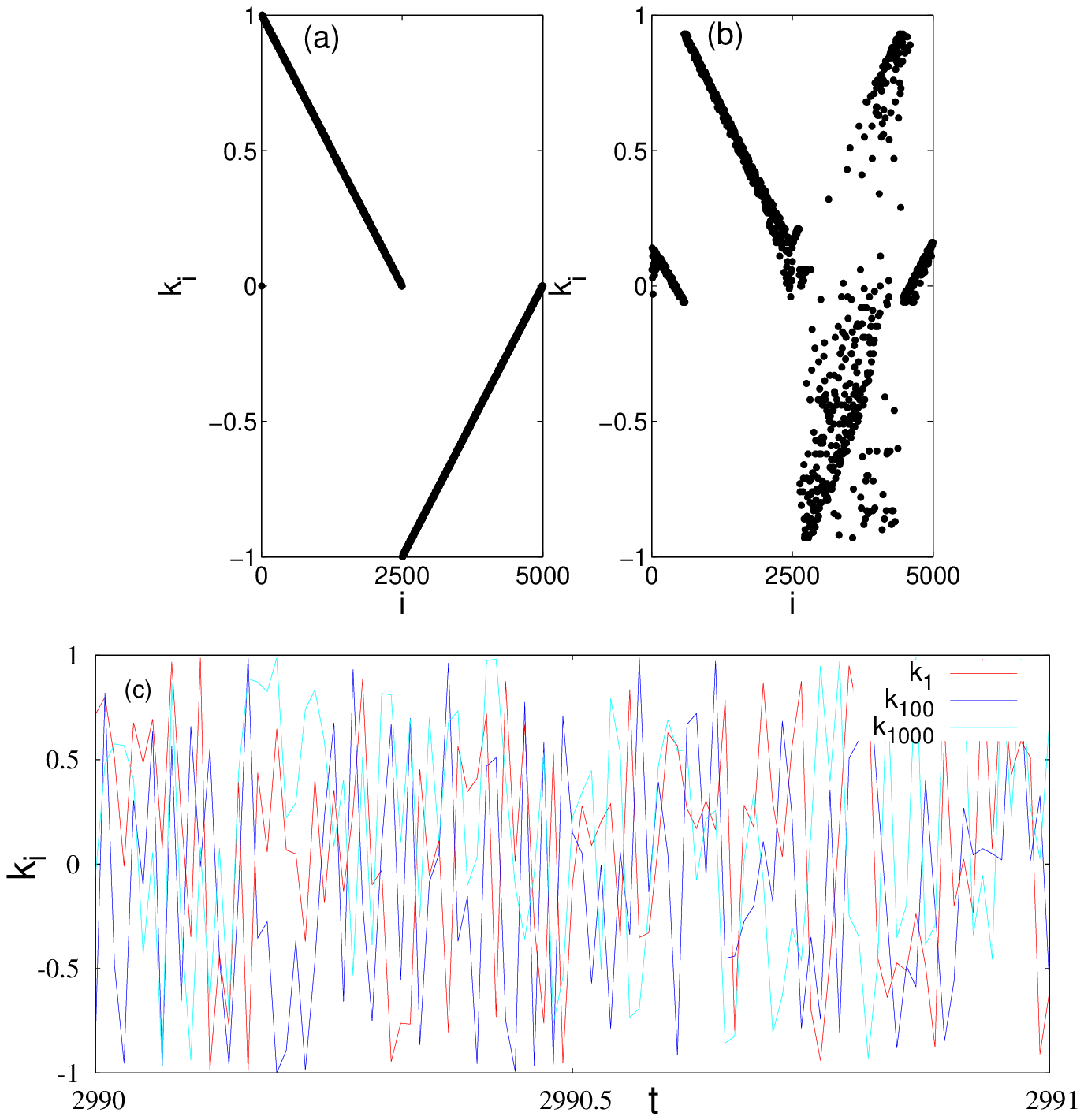}
\caption {Evolution of the coupling weights $k_i$ after leaving out sufficient transients (t=$10^5$). (a) Snapshot of $k_i$  at a particular time in the absence of dynamic coupling ($\eta=0.0$). (b) Snapshot of $k_i$ at a particular time in the presence of dynamic coupling ($\eta=0.05$) and (c) Time evolution of a sample set of three coupling weights $k_i, i=1, 100, 1000$  for a short range of time in the presence of dynamic coupling ($\eta=0.05$) for a random distribution of initial phases.}
\label{figk}
\end{figure}

Further in Fig. \ref{figk} we have plotted the time evolution of coupling weights $k_i$ in the absence and presence of dynamic coupling. Fig. \ref{figk}(a) clearly shows that $k_i(t)=k_i(0)$ in the absence of dynamic coupling ($\eta=0.0$), while Figs. \ref{figk}(b) and \ref{figk}(c) show the random flunctuation of $k_i$'s as a function of time for a random initial distribution of phases. Note that for the case of uniform distribution of initial phases, even in the presence of dynamic coupling ($\eta=0.05$) $k_i$'s continue to remain constant as shown in Fig. \ref{figk}(a).

\subsection{The effect of coupling time scale on the multi-stable state}
\label{Cmulti}
Upon the introduction of coupling plasticity, multi-stable states are found to occur in system (\ref{cho03})  as we have discussed in the previous Section. Now let us see how the time scale of the coupling, denoted by $\eta^{-1}$, affects the multi-stable states. We see that the introduction of $\eta$ causes the occurrence of a two-clustered state, the clusters being in antiphase relationship with each other. It is also obvious that as $\eta$ increases $R_1$ increases while $R_2$ stays close to 1. This essentially means that as $\eta$ increases, the size of one of the clusters grows and as a result, the two cluster state tends to become a single clustered state. So the size difference of the two clustered state $\Delta N$ also grows. In order to explain this behaviour, we move the system to a rotating frame by introducing the transformation $\theta_i \rightarrow \theta^{'}_i=\omega+\theta_i$ and then dropping the primes for convenience. With this transformation, Eq. (\ref{cho03}) becomes
\begin{eqnarray}
\label{asy04}
\dot{\theta_i}&=& \sigma R_1
k_i\sin(\psi_1-\theta_i),\;\;
\dot{k_i}= \eta R_1
\cos(\theta_i-\psi_1)).
\end{eqnarray}
where $R_1$ and $\psi_1$ are as defined in Eq. (\ref{cho01a}).

\begin{figure*}
\begin{center}
\includegraphics[width=17.50cm,height=9.0cm]{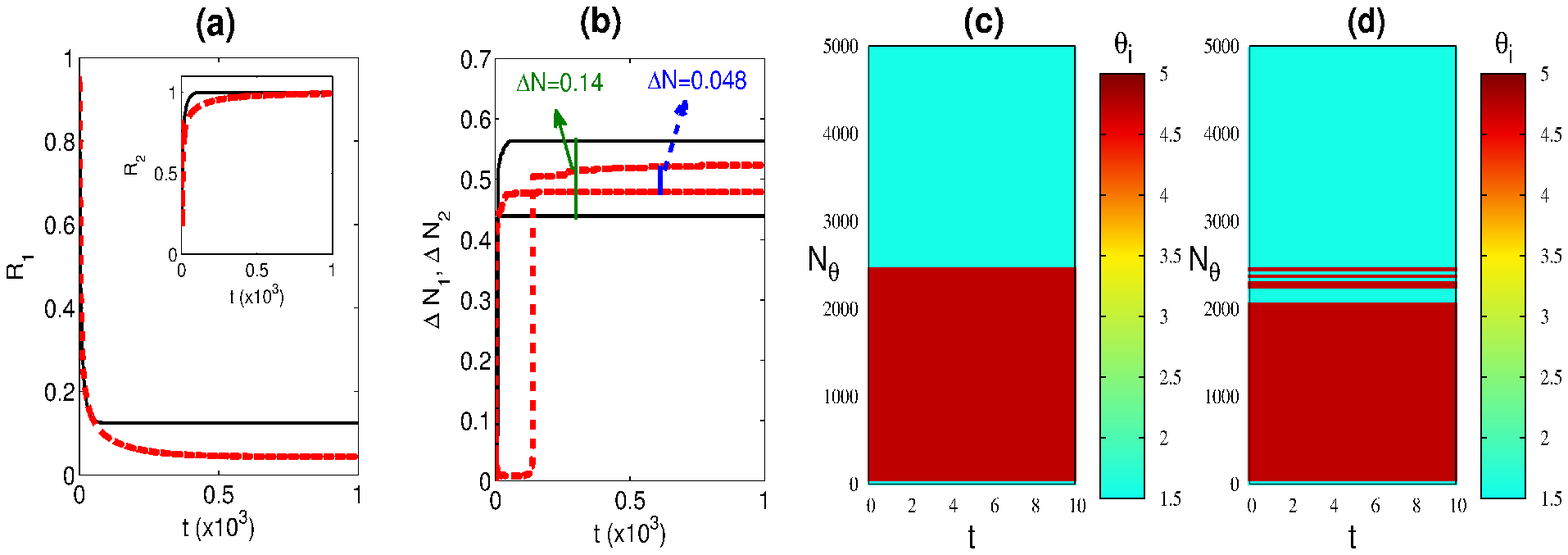}
\caption{ The effect of $\eta$ on the two-cluster state in Eq. (\ref{cho03}). Panel (a): The time evolution of $R_1$ for the two values of $\eta$; $\eta=0.004$ (dashed red line) and $\eta=0.006$ (solid black line). The inset shows the time evolution of $R_2$ for the same two values of $\eta$. Panel (b): The time evolution of the (fraction) number of oscillators in the two clusters, $\Delta N_1=N_1/N$ and $\Delta N_2=N_2/N$ for the two values of $\eta$; $\eta=0.004$ (dashed red line) and $\eta=0.006$ (solid black line). Here the total number of oscillators N=5000.  Panels (c) and (d): Number of oscillators ($N_\theta$) in the two clusters for two values of $\eta$, 1) $\eta=0.04$ (panel(c)) and 2) $\eta=0.06)$ (panel (d)) clearly showing a two cluster state.}
\label{fig4}
\end{center}
\end{figure*}

Now if we choose a random distribution for the initial phases of the oscillators, the presence of $\eta$ stabilizes a two-cluster state in the system; one cluster is locked in $\theta=\pi/2$ and the other is in $\theta=3\pi/2$. This can be clearly seen { \bf from Fig. \ref{fig4} where we have plotted the asymptotic } time evolution of $\theta_i$s for two different values of $\eta$; $\eta=0.004$  and $\eta=0.006$. In panel (a) we have plotted the time evolution of $R_1$ for the two values of $\eta$; $\eta=0.004$ (dashed line) and $\eta=0.006$ (solid line). The inset shows the time evolution of $R_2$ for the same two values of $\eta$ and it is found to be equal to 1 in both the cases. However we see that $R_1$ takes different values for different $\eta$ indicating that the number difference between the two clusters increases with increasing $\eta$ see panels (b),(c) and (d).

In order to visualize this clearly, we have plotted the time evolution of the fraction of oscillators in the two clusters, $\Delta N_1=N_1/N$ and $\Delta N_2=N_2/N$ in panel (b) for the two values of $\eta$; $\eta=0.004$ (dashed line) and $\eta=0.006$ (solid line). Here $N=5000$ is the total number of oscillators. Since there are only two clusters in the system of sizes $N_1$ and $N_2$, we choose a random oscillator (say, the first oscillator whose phase is denoted by $\theta_1$), and compare the phases of all the other oscillators in the system with that of oscillator 1. Now, if the phase of any oscillator is equal to $\theta_1$, then that oscillator is in the same cluster as that of the 1st oscillator and let us assume that this cluster has $N_1$ oscillators. On the other hand, if the phase of the oscillator is not equal to $\theta_1$, then that oscillator is in the other cluster whose size is $N_2$. This is how we count the number of oscillators in the two clusters numerically and have plotted the corresponding fractions $\Delta N_1$ and $\Delta N_2$  in panel (b). In panels (c) and (d) we have shown the number of oscillators in each of the two clusters $N_{\theta}$ as a function of time in the asymptotic regime for  $\eta=0.004$  and $\eta=0.006$, respectively. The number difference between the two clusters increases for increasing $\eta$ and the value of $\Delta N$ matches the value of $R_1$ for the corresponding $\eta$, in panel (a). Thus the coupling time scale affects the size of the clusters in the two-cluster state.

\section{The effect of phase asymmetry in coupling plasticity}
\label {phase lag}
\begin{figure*}
\begin{center}
\includegraphics[width=14.5cm,height=7.5cm]{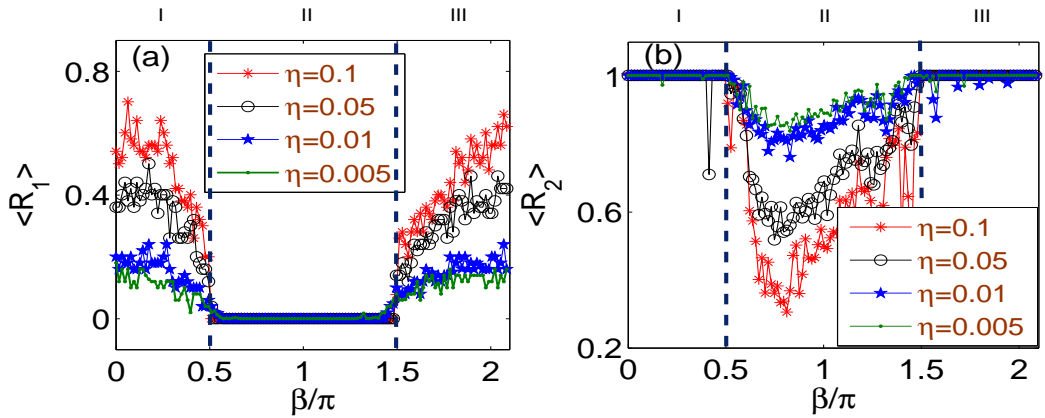}
\caption{ Time averaged value of order parameters, $<R_1>$ (panel (a)) and $<R_2>$ (panel (b)), are plotted against $\beta$ for different values of $\eta$. Regions I, II and III represent two clustered, desynchronization and two clustered states, respectively.}
\label{fig5}
\end{center}
\end{figure*}
\begin{figure*}
\begin{center}
\includegraphics[]{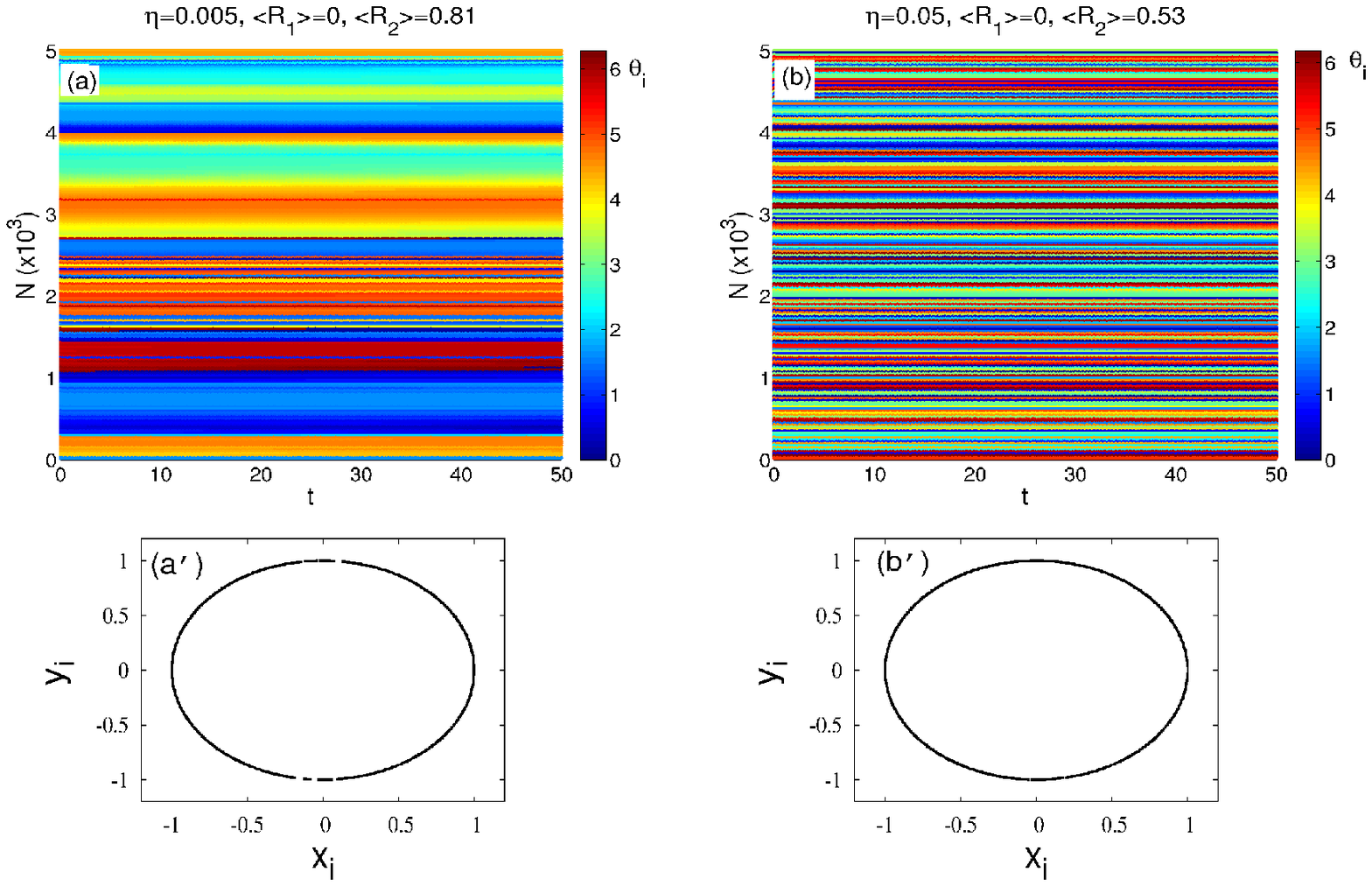}
\caption{ The asymptotic  time evolution of the oscillator phases plotted for $\eta=0.005$ (a), and  $\eta=0.05$ (b).  Phase desynchronized nature of these states are clearly evident from the uniform distribution of phases in the range [0, $2\pi$] in the phase portrait (panels (a'), (b'), respectively) shown below their corresponding panels. The other parameters are: $\beta=3.0$, $\sigma=1.0$, and $N=5000$.}
\label{figphase}
\end{center}
\end{figure*}
We now introduce phase asymmetry parameters $\alpha$ and $\beta$ in (\ref{cho03}).  The transmission interlude or delay of the coupling can be represented by the phase difference $\alpha$.  On the other hand, the characteristic of plasticity can be continuously changed or controlled by varying a second asymmetry parameter $\beta$. Thus this parameter $\beta$, which is the plasticity delay, enables one to investigate the coevolving dynamics \cite{Aoki:11}.
With the introduction of the asymmetry parameters, the coupling function  $f(\theta)$ and the plasticity function  $g(\theta)$ in (\ref{cho01})  and (\ref{cho02})  become $f(\theta)=\sin(\theta+\alpha)$ and $g(\theta)=\cos(\theta+\beta)$. With these asymmetry parameters, the evolution equations of system (\ref{cho01})  and (\ref{cho02}) become
 \begin{eqnarray}
\label{asy01a}
\dot{\theta_i}&=& \omega- \frac{\sigma}{N}\sum_{j=1}^{N}
k_i\sin(\theta_i-\theta_j+\alpha),\\
\dot{k_i}&=& \frac{\eta}{N}\sum_{j=1}^{N}
\cos(\theta_i-\theta_j+\beta),\; i=1,2, \ldots, N,
\label{asy01b}
\end{eqnarray}
 and we choose $0 \leq \alpha \leq \pi/2$ and $0 \leq \beta \leq 2\pi$.
 
\begin{figure*}
\begin{center}
\includegraphics[width=14.50cm]{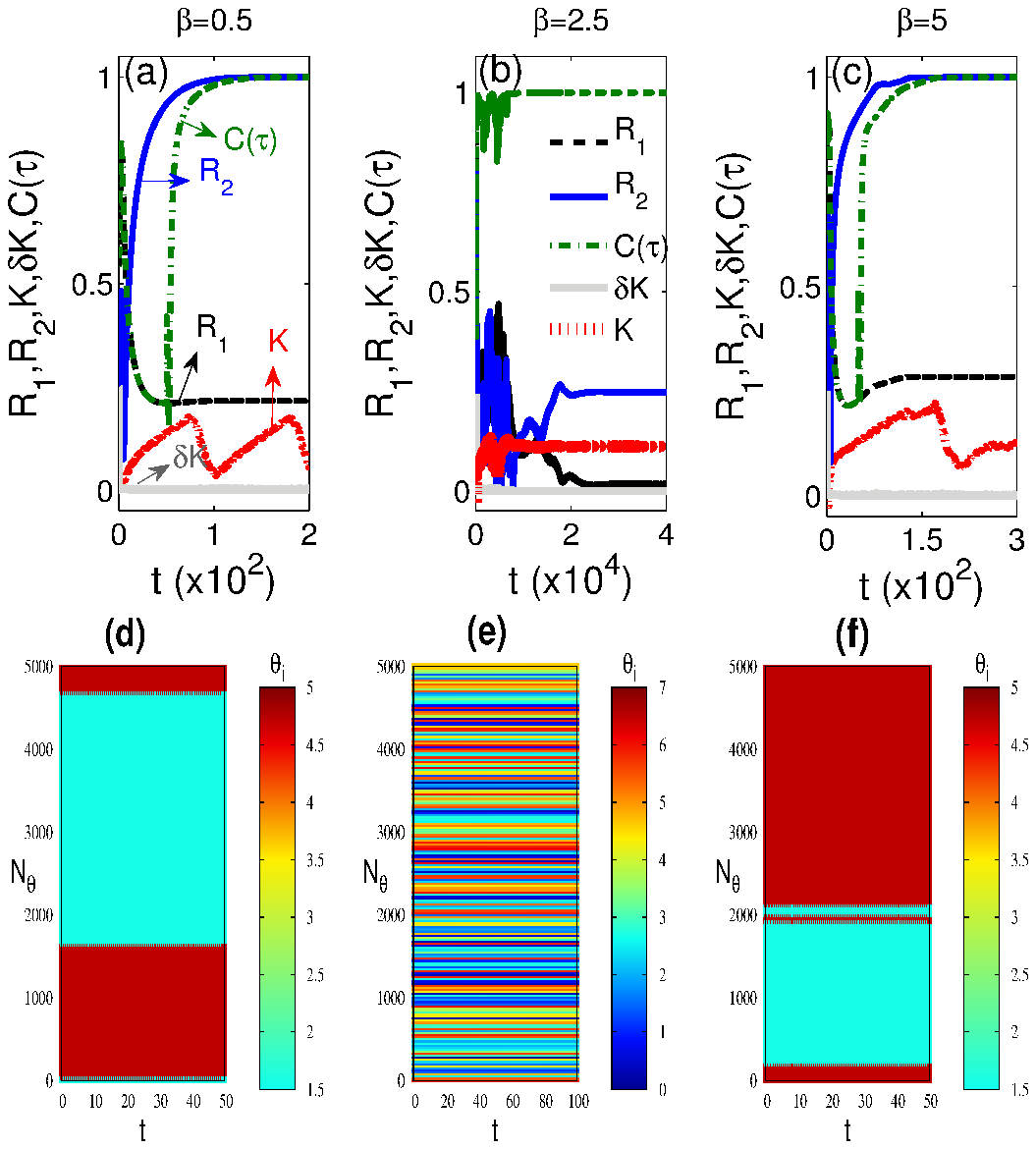}
\caption{ The effect of $\beta$ on regions I, II and III corresponding to Fig. \ref{fig5}. Panels (a)-(c): the time evolution of $R_1$ (dashed black line), $R_2$ (solid  blue line),  $K$ (dotted red line) and $\delta K$ (solid grey line) and the autocorrelation $C(\tau)$ (dot-dashed green line) as shown in the legend of panel (b). Panels (d)-(f) correspond the asymptotic time evolution of the oscillator phases (Oscillator number is designated as $N_\theta$).  The chosen $\beta$ values are as follows : 1) For panels (a), (d) : $\beta=0.5$,  (2) For panels (b), (e) : $\beta= 2.5$  and (3) For panels (c), (f)$\beta= 5.0$ . The other parameters are: $\eta=0.05$, $\sigma=1.0$, and $N=5000$.}
\label{fig6}
\end{center}
\end{figure*}

In order to study the influence of the plasticity asymmetry parameter $\beta$ on the occurrence of multi-stable states in the absence of the coupling asymmetry parameter $\alpha=0$, we have plotted the time averaged value of the order parameters $R_k$, k=1,2, given as
\begin{eqnarray}
\label{cho01b}
<R_k>=\frac{1}{T}\int_{0}^{T}R_kdt, \;k=1,2,
\end{eqnarray}
against $\beta$ for different values of $\eta$ in Fig. \ref{fig5}. In panel (a) we have plotted $<R_1>$ and we see that as $\beta$ increases, the two-clustered state exists until $\beta$ approaches a critical value of $\sim\pi/2$. At the same time we see that $R_1$ decreases with increasing $\beta$. In this window, we see that the corresponding value of $<R_2>$ (plotted in panel (b)) stays at 1 implying that the size difference $\Delta N$ between the two clusters decreases. After the transition point $\beta=\pi/2$ the two-cluster state loses its stability and only the desynchronized state is stable. This can be confirmed by the decreasing $<R_2>$ after the transition point $\beta=\pi/2$. When $\beta$ takes a value in the window $\pi/2<\beta<3\pi/2$ only the desynchronized state is stable; this is evident from the values of $<R_1>$ which is zero and $<R_2>$ which takes a value less than 1. When $\beta>3\pi/2$, the two clustered state becomes stable again leading to the occurrence of multi-stability. In this window ($3\pi/2<\beta<2\pi$) as $\beta$ increases $<R_1>$ increases and $<R_2>$ takes a value close to 1 indicating that the size difference between the two clusters $\Delta N$ increases. Thus we find that in the absence of $\alpha$, increasing $\beta$ for a given $\eta$ causes the system to go from a two-cluster state (region I) to desynchronization (region II) and then again to a two-cluster state (region III). It is also evident that the occurrence of multi-stable states due to the asymmetry parameter $\beta$ is unaffected by $\eta$. In Fig. \ref{figphase} we have plotted the time evolution of phases for two different values of $\eta$, namely $\eta=0.005$ (in panel (a)) and $\eta=0.05$ (in panel (b)) for a fixed value of $\beta=3.0$.  In both the cases the time averaged order parameters $<R_1>$ takes the value zero and $<R_2>$ takes some finite value other than zero, $<R_2>$=0.81 ($\eta=0.005$) and $<R_2>=0.53$ $(\eta=0.005$), respectively. These figures clearly show the phase desynchronized nature of the oscillators, irrespective of the non zero values of $<R_2>$. Similarly, the corresponding phase portraits in panels (a) and (b) respectively, for both the values of $\eta$ clearly show the uniformly distributed nature of phases between 0 and 2$\pi$. We have also confirmed similar features for $\eta=0.01$ and $\eta=0.1$ for the same value of $\beta$. One may also observe that the autocorrelation function, $C(\tau)$,  asymptotically takes a unit value, confirming a fixed phase relationship between the phases, and excluding any chaotic behaviour (see below). Due to this fact, we call the corresponding state (region II) as a desynchronized state even though $<R_2>$ takes some finite value other than zero.

In order to explain the above mentioned states that occur due to the effect of $\beta$ more clearly, let us refer to Fig. \ref{fig6}. In panels (a)-(c) we have plotted the time evolution of $R_1$ (dashed black line), $R_2$ (solid blue line), $K$ (dotted red line),  $\delta K$ (solid grey line) and the autocorrelation $C(\tau)$ (dot-dashed green line). The asymptotic time evolution of the oscillator phases are plotted in panels (d)-(f). The panels (a) and (d) correspond to $\beta=0.5$, panels (b) and (e) correspond to $\beta=2.5$ and panels (c),  and (f) correspond to $\beta=5.0$. When $\beta=0.5$ the two cluster state is stable; this state corresponds to region I in Fig. \ref{fig5} and the two clusters are of different sizes (panel (d)). In this state $R_1>0$ and $R_2=1.0$. When $\beta=2.5$, the two cluster state loses its stability leaving behind only the desynchronized state; this state corresponds to region II in Fig. \ref{fig5}. Since the oscillators are desynchronized and since there are no clusters in the system, we cannot define $N_1$ and $N_2$ in this state as shown in panel (e). In this state $R_1\sim0$ and $R_2\sim0.25$. When $\beta=5.0$, near the II-III transition point, the two cluster state becomes stable again and this state corresponds to region III in Fig. \ref{fig5}.  Note that in all these cases the autocorrelation function $C(\tau)$ takes a unit value asymptotically, confirming the fixed phase relationship nature of the oscillators.

\begin{figure*}
\begin{center}
\includegraphics[width=8.50cm]{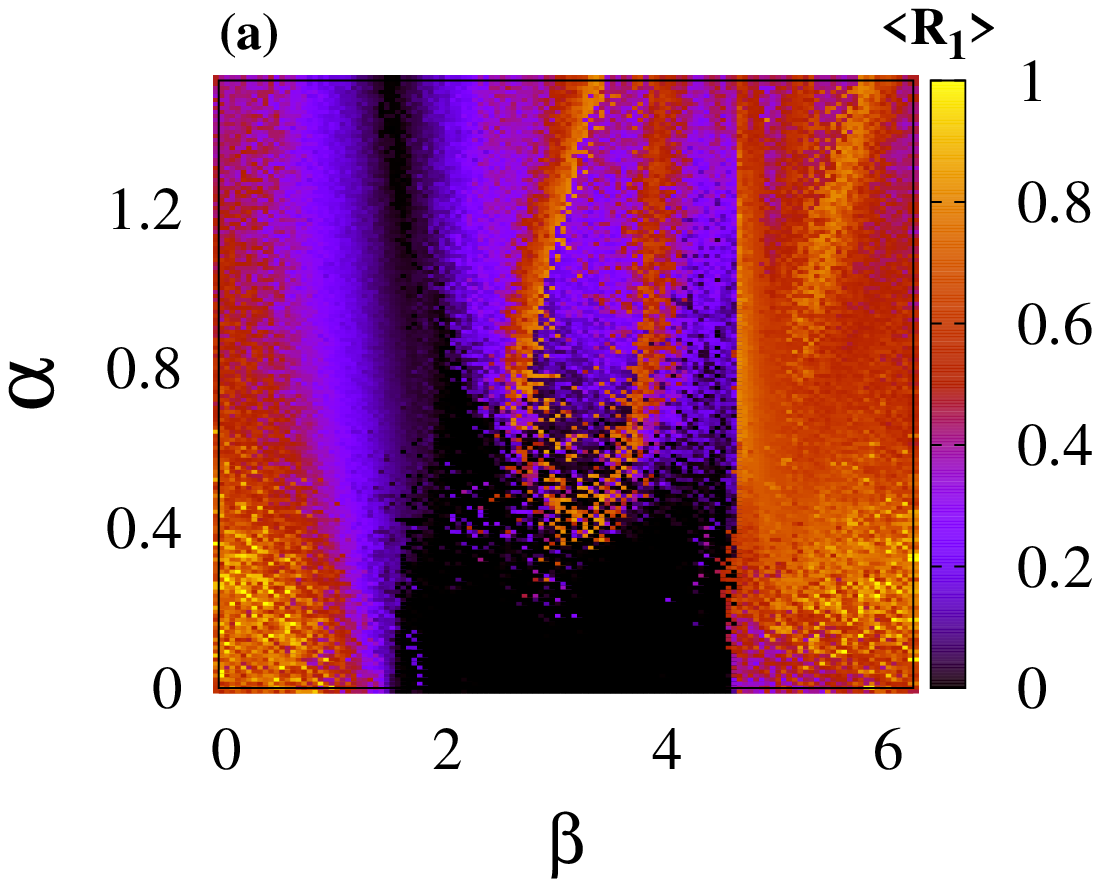}
\includegraphics[width=8.50cm]{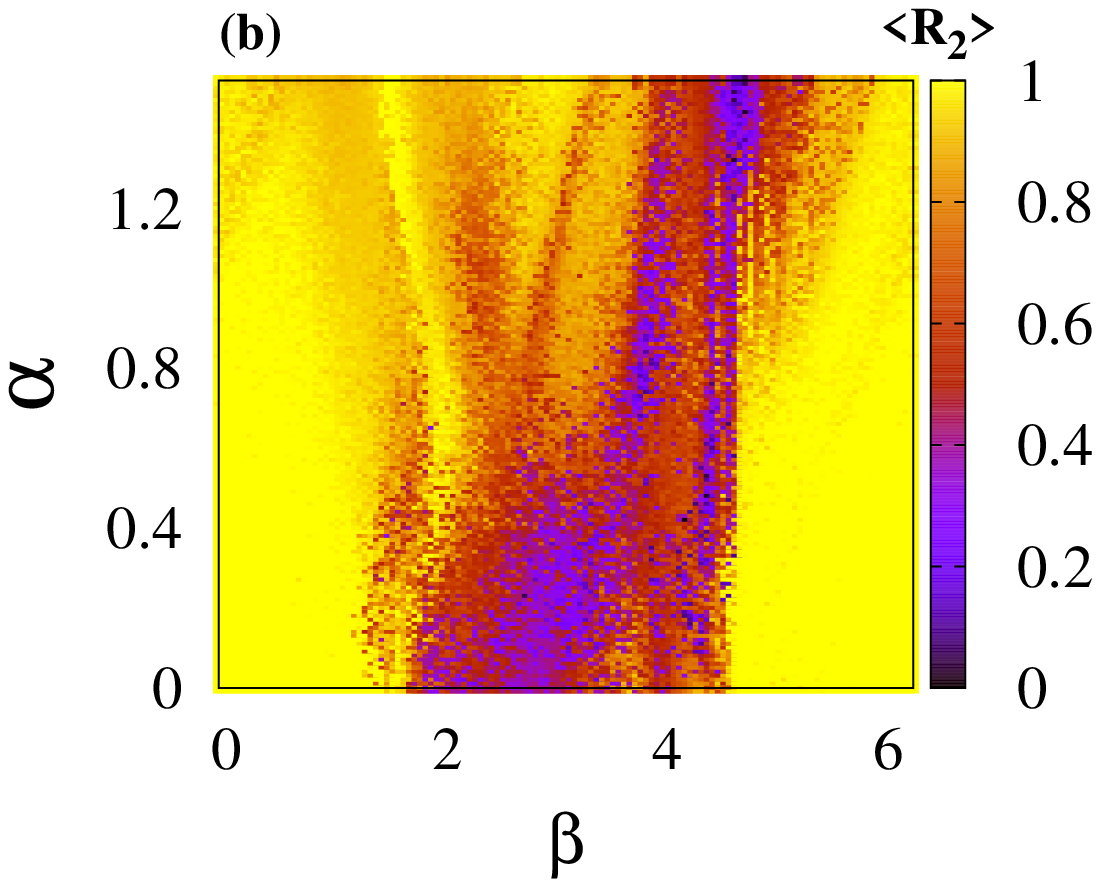}
\caption{Two parameter phase diagram showing the strengths of $<R_1>$ (panel (a) and $<R_2>$ (panel (b)) for varying $\alpha$ and $\beta$. The other parameters in Eqs. (\ref{asy01a})-(\ref{asy01b}) are: $\omega=1.0$, $\eta=0.1$, $\sigma=1.0$, and $N=1000$.}
\label{fig7}
\end{center}
\end{figure*}

Now, in order to study the effect of the coupling asymmetry parameter $\beta$ in the presence of $\alpha$ we have plotted the strengths of $<R_1>$ (panel (a)) and $<R_2>$ (panel (b)), by varying both $\alpha$ and $\beta$ in Fig. \ref{fig7}. While varying $\alpha$ from 0 to $\pi/2$, we see that the regions I and III (corresponding to Fig. \ref{fig5}) shrink while region II expands. Further, as $\alpha$ increases we find that the size of the clusters in regions I and III keeps varying which is evident from the varying strengths (colors) of $<R_1>$ in panel (a) while the strength of $<R_2>$ remains the same in panel (b). Thus we find that the coupling asymmetry affects the influence of $\beta$ on the two cluster and desynchronization states (regions I, II and III).

\section{Stability of the  desynchronized state and the two cluster state}
\label {ana}
In this Section we wish to investigate analytically the linear stability of the desynchronized state and the two clustered synchronized state. Numerically, we find that the oscillator phases maintain a fixed relationship among themselves and the coupling weights remain almost stable when $\eta\rightarrow 0$ (see insets in Fig. 2 (b), (e)). Hence in the limit $\eta\rightarrow 0$, the coupling weights can be regarded as invariant and remain at a constant value. Under this condition, the dynamics of system (\ref{asy01a})-(\ref{asy01b}) with the asymmetry parameters $\alpha$ and $\beta$ is given by
 \begin{eqnarray}
\label{ana01}
\dot{\theta_i}&=& \omega- \frac{\sigma}{N}\sum_{j=1}^{N}
k_i\sin(\theta_{ij}+\alpha),
\end{eqnarray}
where $\theta_{ij}=\theta_i-\theta_j$. Since the coupling weights are fixed, they satisfy the following relation
\begin{eqnarray}
\label{ana02}
k_i=sgn(\frac{1}{N}\sum_{j=1}^{N}
\cos(\theta_{ij}+\beta)).
\end{eqnarray}
Now we consider two different state configurations separately, (i) Uniform distribution of phases and (ii) a two cluster state and analyze the linear stability of the underlying states.

\subsection{Uniformly distributed phases}
\label{uniform_analy}
Now let us assume that the phases are uniformly distributed in $[0,A]$, where $A$ is an arbitrary constant. So the phase relationship between the $i$th and the $j$th oscillators is given by
\begin{eqnarray}
\label{ana03}
\theta_{ij}=\frac{A(j-i)}{N}.
\end{eqnarray}
\begin{figure}
\begin{center}
\includegraphics[width=8.50cm]{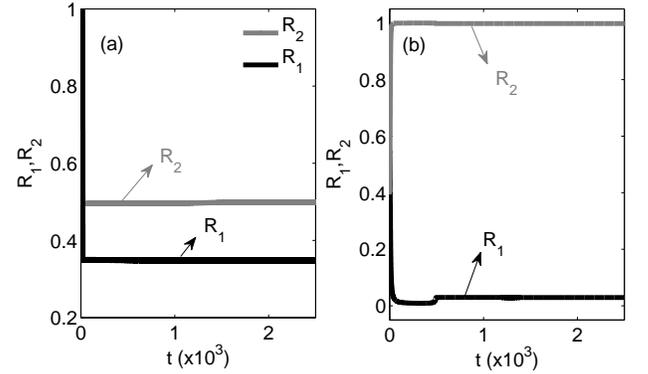}
\caption{ The time evolution of the order parameters $R_1$ (solid black) and $R_2$ (solid gray) for the asymmetry parameters $\alpha$=1.5 and $\beta$=1.0 for two different distributions of initial phases. Panel (a) corresponds to a random distribution of initial phases between 0 and $\pi$ which shows the existence of a partially synchronized state as in Fig.\ref{fig7}. On the other hand panel (b) shows a two cluster state ($R_2=1$) for the same set of asymmetry parameters but with the initial phases chosen to be in a two cluster state.}
\label{fig11}
\end{center}
\end{figure}

When $N$ approaches the limit $N\rightarrow \infty$, Eq. (\ref{ana02}) becomes
\begin{eqnarray}
\label{ana05}
k_i&=&\mbox{sgn}(\frac{1}{A}\int_{0}^{A}
\cos(\theta+\beta)d\theta),\nonumber\\
&=&\mbox{sgn}(\frac{1}{A}(
\sin(A+\beta)-\sin(\beta))).
\end{eqnarray}
For $A=2\pi$, that is when the phases are uniformly distributed in $[0,2\pi]$, the integral in Eq. (\ref{ana05}) vanishes and hence $k_i=k_i(0)$ (when the plasticity is absent). 

In this case, let us rewrite Eq. (\ref{ana01}) as 
\begin{eqnarray}
\label{ana01a}
\dot{\phi_i}&=& \omega-\Omega- \frac{\sigma}{N}\sum_{j=1}^{N}
k_i\sin(\phi_{ij}+\alpha),
\end{eqnarray}
where $\phi_i=\theta_i-\Omega t$ and $\Omega$ is the collective frequency of oscillation of the population after the system approaches a stationary state. Using the order parameter relation (\ref{cho01a}) into Eq. (\ref{ana01a}) we get
\begin{eqnarray}
\label{ana01b}
\dot{\phi_i}&=& \omega-\Omega- \sigma R_1
k_i\sin(\phi_{i}-\Phi+\alpha),
\end{eqnarray}
where $\Phi=\psi_1-\Omega t$  and $\psi_1$ is as defined in Eq. (\ref{cho01a}). Now, we consider the $N \rightarrow \infty$ thermodynamic limit. In this case the order parameter can be written as \cite{Strog:20,Gab:08}
\begin{eqnarray}
\label{ana01c}
R_1e^{i\Phi}=\int_{-\infty}^{\infty}dk\int_{0}^{2\pi}e^{i\phi}n(\phi;k) d\phi,
\end{eqnarray}
where $n(\phi;k)$ is the density distribution of the individual phases $\phi$ with coupling $k$. In the stationary state, $R_1$ and $\Phi$ are time independent constants and hence Eq. (\ref{ana01c}) becomes
\begin{eqnarray}
\label{ana01d}
R_1=\int_{-\infty}^{\infty}dk\int_{0}^{2\pi}e^{i\phi}n(\phi;k) d\phi,
\end{eqnarray}
where we have taken $\Phi=\Phi(0)=0$ without loss of generality. When $|\omega-\Omega| \leq k_i\sigma R_1$, Eq. (\ref{ana01a}) has a stable fixed point at $\phi_i=sin^{-1}((\omega-\Omega)/(k_i\sigma R_1))$ with $-\pi/2 \leq \phi_i \leq \pi/2$. Now one can consider the relation 
\begin{eqnarray}
\label{ana01e}
n(\phi;k) d\phi dk  =k\sigma R_1 \cos \phi d\phi dk=h(k) d\omega dk
\end{eqnarray}
into Eq. (\ref{ana01d}) to arrive at the self-consistency equation
\begin{eqnarray}
\label{ana01f}
1  =\sigma  \int_{-\infty}^{\infty}k h(k) dk\int_{-\pi/2}^{\pi/2}\cos^2\phi d\phi,
\end{eqnarray}
and 
\begin{eqnarray}
\label{ana01g}
0  =\sigma  \int_{-\infty}^{\infty}k h(k) dk\int_{-\pi/2}^{\pi/2}\cos\phi \sin\phi d\phi.
\end{eqnarray}
From the above relations (\ref{ana01f}) and (\ref{ana01g}), the synchronization threshold is defined by the condition that the average coupling strength
\begin{equation} \label{kc}
<k>  =\int_{-\infty}^{\infty} k h(k) dk  = \frac{2}{\sigma \pi}= k_c,
\end{equation}
as shown in ref.\cite{Gab:08} and the synchronized state exists when  $<k>$ $\ge$ $k_c$.
Here in our case in the absence of coupling plasticity, $k_i=k_i(0)$ and these are distributed uniformly between -1 and +1. Consequently $h(k)=1$ and 
\begin{equation} \label{kc}
<k> = \int_{-1}^{+1} k dk  = 0,
\end{equation} 
that is the average coupling strength vanishes and therefore the synchronization state does not exist as it violates the threshold condition (\ref{kc}) and the phases remain desynchronized showing the stable nature of the desynchronized state \cite{Gab:08}.

\subsection{Two cluster state}
On the other hand, in the presence of coupling plasticity, in order to study the stability of the two cluster state, let us assume that the the size of the two clusters are $N_1$ and $N_2$ whose phases are $\theta_1$ and $\theta_2$, respectively. In the two cluster state, Eq. (\ref{ana01}) is written as
\begin{eqnarray}
\label{ana06}
\dot{\theta_1}&=& -\frac{\sigma N_1}{N}k_1\sin(\alpha)-\frac{\sigma N_2}{N}k_1\sin(\Delta \theta+\alpha),\\
\dot{\theta_2}&=& -\frac{\sigma N_2}{N}k_2\sin(\alpha)+\frac{\sigma N_1}{N}k_2\sin(\Delta \theta-\alpha)
\end{eqnarray}
and the coupling weights (Eq. (\ref{ana02})) become
\begin{eqnarray}
\label{ana07}
k_1&=&\mbox{sgn}(\frac{N_1}{N}\cos(\beta)+\frac{N_2}{N}\cos(\Delta \theta+\beta))\nonumber\\
k_2&=&\mbox{sgn}(\frac{N_2}{N}\cos(\beta)+\frac{N_1}{N}\cos(\Delta \theta-\beta)).
\end{eqnarray}
The phase difference between $\theta_1$ and $\theta_2$ can be written as
\begin{eqnarray}
\label{ana08}
\dot{\Delta\theta}&=& -(\frac{\sigma N_1}{N}k_1-\frac{\sigma N_2}{N}k_2)\sin(\alpha)-\frac{\sigma N_2}{N}k_1\sin(\Delta \theta+\alpha)
\nonumber\\&&-\frac{\sigma N_1}{N}k_2\sin(\Delta \theta-\alpha).
\end{eqnarray}
Now, the stability condtion for the two cluster state is determined from  (\ref{ana08}) as
\begin{eqnarray}
\label{ana09}
0&>&-\sigma (\frac{ N_1}{N}k_2\cos(\Delta \theta-\alpha)+\frac{ N_2}{N}k_1 \cos(\Delta \theta+\alpha)).
\end{eqnarray}
On using (\ref{ana07}) in (\ref{ana08}), we obtain
\begin{eqnarray}
\label{ana10}
0&>&-\sigma (\frac{N_1}{N}\mbox{sgn}(\frac{N_2}{N}\cos(\beta)+\frac{N_1}{N}\cos(\Delta \theta+\beta))\nonumber\\
&&\quad\times\cos(\Delta \theta-\alpha)
+\frac{ N_2}{N} \mbox{sgn}(\frac{N_1}{N}\cos(\beta)\nonumber\\
&&\quad+\frac{N_2}{N}\cos(\Delta \theta-\beta))\cos(\Delta \theta+\alpha))
\end{eqnarray}
Numerically we find that, in the two cluster state, the phase difference between the clusters is $\pi$ (for instance, see Figs. 4 and 6). Hence, for the case $\theta_2=\theta_1+\pi$ the stability condition becomes
\begin{eqnarray}
\label{ana11}
0&>&-\sigma \Delta N \mbox{sgn}(\Delta N\cos(\beta))\cos(\alpha).
\end{eqnarray}
This condition represents the stability of the two cluster state. When $\alpha \in [0,\pi/2]$ and $\sigma>0$, only $\beta$ contributes to the sign of the right hand side in the above equation. Hence the two cluster state is stable in the windows $\beta \in [0,\pi/2]$ and $[3\pi/2, 2\pi]$. The two cluster state loses stability and only the desynchronized state becomes stable in the window $\pi/2<\beta<3\pi/2$. For increasing the plasticity asymmetry parameter $\beta$ the two cluster-desynchronization-two cluster transition occurs (as shown in Fig. \ref{fig5}).
We see that our numerical observations (Fig. \ref{fig7}) are in good agreement with the analytical results in the region of phase asymmetry $\alpha$, $0<\alpha<\pi/4$ approximately, beyond which there arises disagreement where a nonlinear stability theory will be required. In order to explain the above disagreement, we have performed numerical simulations of Eq. (\ref{ana01}) obeying Eqs. (\ref{ana06})-(\ref{ana07}), that is initial phases are chosen to be in a two cluster state. Fig. \ref{fig11}(b) shows the existence of a two cluster state of the system in the asymmetry region $\alpha$ = 1.5 and $\beta$ = 1.0, which does not exist in Fig. \ref{fig7} as well as in panel (a) of Fig. \ref{fig11}, where we have chosen random initial phases. Thus we find that the two cluster state arises only for the special choice of initial conditions near to the above state, while for an arbitrary initial distribution of phases one obtains a partially synchronized state, indicating the multistable nature of the underlying system.

\section{The case of Nonidentical oscillators}
\label{nio}

In the previous sections, we have studied the effect of plastic coupling between the group of identical oscillators($\omega$ = constant). In the present section we briefly consider the case of nonidentical oscillators  whose frequencies ($\omega_{i}$) are distributed in Lorentzian form given by
\begin{eqnarray}
\label{cho10}
g(\omega)&=&\frac{\gamma}{\pi}\bigg[(\omega-\omega_{0})^2+\gamma\bigg]^{-1},
\end{eqnarray}
where $\gamma$ is the half width at half maximum and $\omega_{0}$ is the central frequency. Hence the equation of motion for the $N$ nonidentical oscillators can be written as
 \begin{eqnarray}
\label{cho11}
\dot{\theta_i}&=& \omega_i - \frac{\sigma}{N}\sum_{j=1}^{N}
k_i\sin(\theta_i-\theta_j),\nonumber\\
\dot{k_i}&=& \frac{\eta}{N}\sum_{j=1}^{N}
\cos(\theta_i-\theta_j),\; i=1,2, \ldots, N.
\end{eqnarray}
\begin{figure*}
\begin{center}
\includegraphics[width=14.5cm,height=10.5cm]{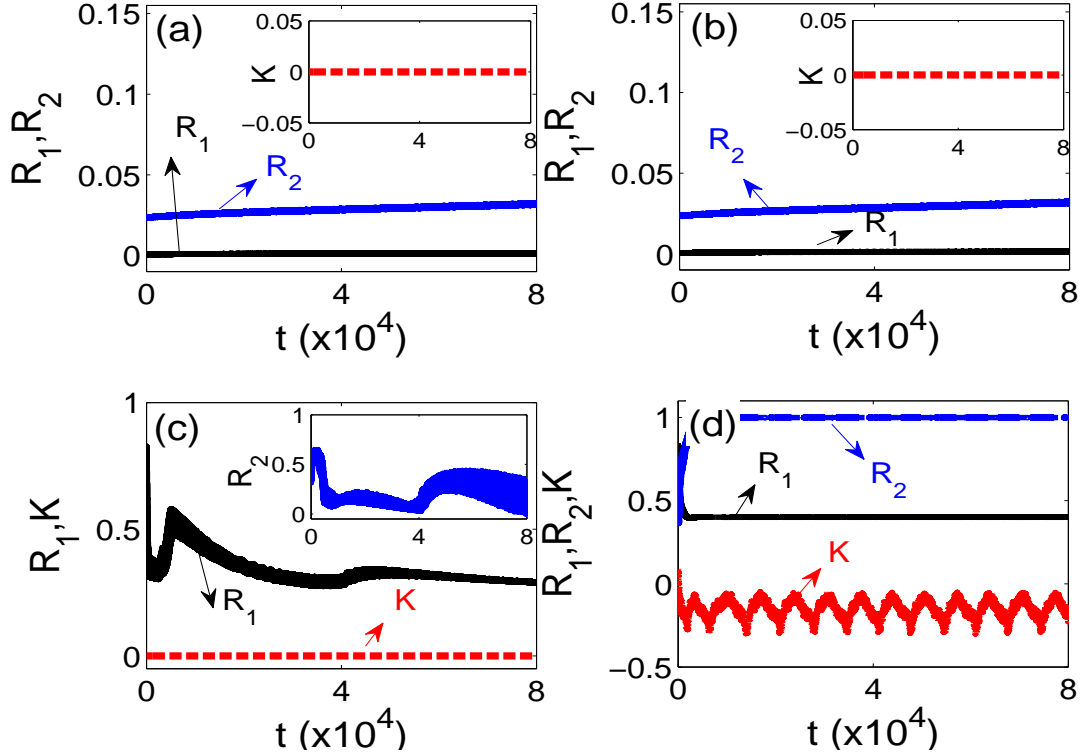}
\caption{ \bf The effect of coupling plasticity for the case of nonidentical oscillators ($\gamma=0.1\times 10^{-5}$) are shown in this figure for two initial configurations. (i) Panels (a)-(b) correspond to uniform distribution in the range [0,2$\pi$]. (ii) Panels (c)-(d) correspond to random distribution in the range [0,$\pi$]. In panels (a) and (c) we have plotted the order parameters $(R_1,R_2)$ and coupling strength $K$ which clearly show that the oscillators are desynchronized in the absence of adaptive coupling for both the initial configurations. Panels (b) and (d) show the onset of multistable state in the system due to the presence of plastic coupling between the nonidentical oscillators. }
\label{fignon}
\end{center}
\end{figure*}
Fig. \ref{fignon} shows the presence of a multistable state for the case of nonidentical oscillators with a very low half width ($\gamma=0.1 \times 10^{-5}$) for $N=5000$ oscillators and $\sigma=1.0$. In the absence of adaptive coupling the system is asymptotically stable in the desynchronized state ($R_1\sim0$) which is evident in panels (a) and (c). However, in the presence of adaptive coupling ($\eta=0.05$), for the uniform distribution of initial phases the system remains in desynchronized state ($R_1\sim0$) see panel (b), whereas for a random distribution the system sets into a different multicluster state ($R_1>0$ and $R_2\sim1$). In Fig. \ref{fig8}, for a higher value of half width ($\gamma=0.1\times10^{-3}$) we find that the multicluster state loses its stability. The full details of the classification of this multicluster state for the case of nonidentical oscillators will be presented elsewhere.
\begin{figure*}
\begin{center}
\includegraphics[width=14.50cm,height=10.50cm]{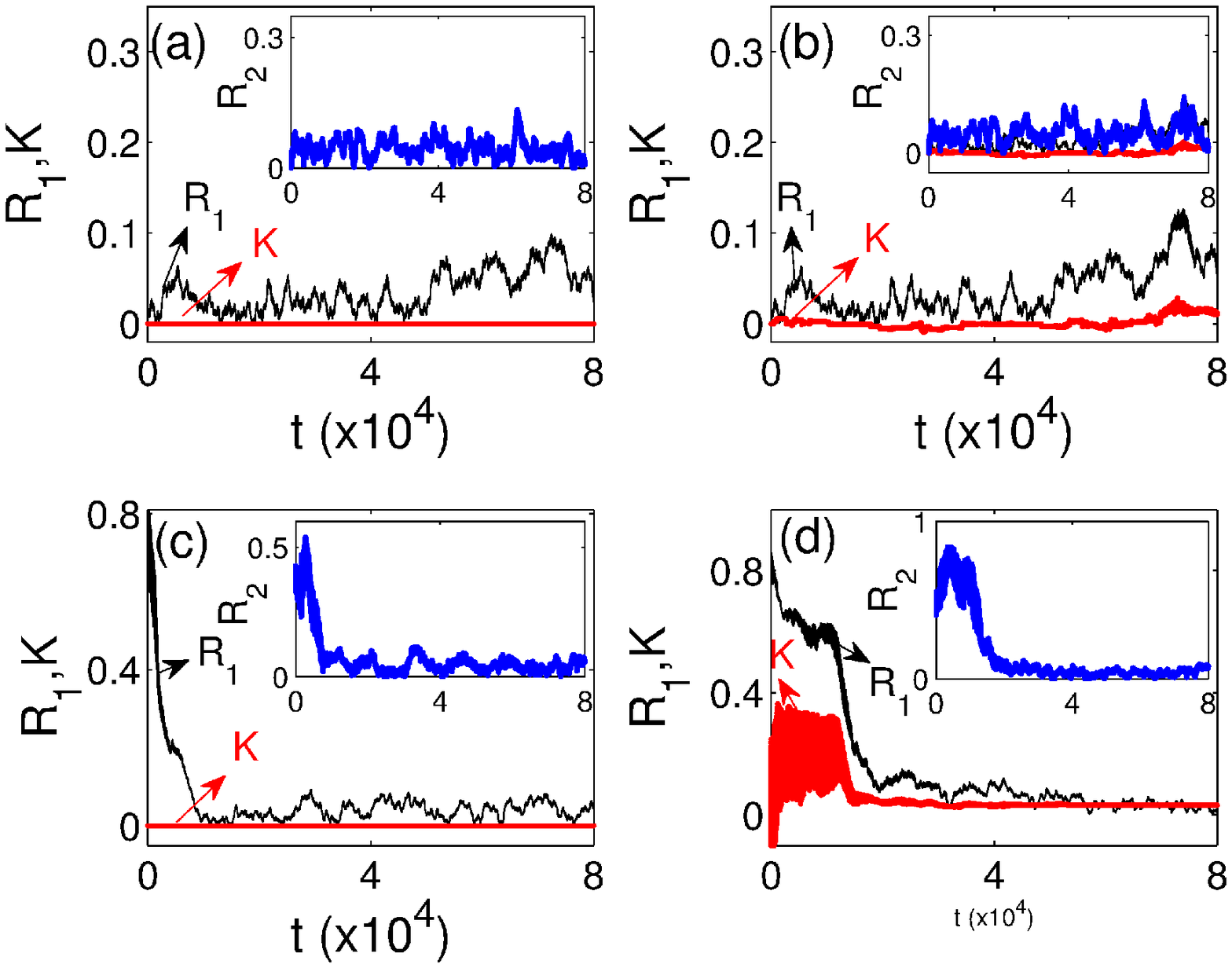}
\caption{The multicluster state loses its stability for $\gamma=0.1\times10^{-3}$ is shown in this figure for two initial configurations similar to Fig.\ref{fignon}. (i) Panels (a)-(b) correspond to uniform distribution in the range [0,2$\pi$]. (ii) Panels (c)-(d) correspond to random distribution in the range [0,$\pi$]. In panels (a) and (c) we have plotted the order parameters $R_1$,$R_2$ (inset) and coupling strength $K$ which clearly show that the oscillators are desynchronized in the absence of adaptive coupling for both the initial configurations. Panels (b) and (d) also show the desynchronized state of the system in the presence ($\eta=0.05$) of plastic coupling between the nonidentical oscillators.}
\label{fig8}
\end{center}
\end{figure*}

\section{Conclusion}
\label{conc}
In this paper, we have demonstrated the occurrence of multi-stable states in a system of phase oscillators that are dynamically coupled. We find that the presence of coupling plasticity induces the occurrence of such multi-stable states; that is, the existence of a desynchronized state and a two-clustered state where the clusters are in anti-phase relationship with each other. The multi-stable states occur for randomly distributed initial phases while for uniformly distributed initial phases only the desynchronization state exists. We also find that the phase relationship between the oscillators is asymptotically stable irrespective of whether there is synchronization or desynchronization in the system. 

We find that the effect of coupling time scale ($\eta^{-1}$) is not only to introduce a two clustered state but also to change the number of oscillators in the two clusters. More precisely, we see that the difference between the number of oscillators in the two clusters increases with increasing $\eta$. In short, the coupling time scale is found to affect the size of the clusters in the two cluster state. 

We have also investigated the effect of coupling asymmetry and plasticity asymmetry on the multi-stable states. We find that, in the absence of coupling asymmetry, increasing plasticity asymmetry takes the system to transit from a multi-stable state through a desynchronized state to a multi-stable transition. If the coupling asymmetry is present, we find that the regions corresponding to two cluster states shrink and the region corresponding to desynchronization state expands.

In our model, for a uniform distribution of initial phases, the desynchronized state is always stable. For random initial conditions, the system goes from a two cluster synchronization state (the desynchronization state exists also in this state) to a desynchronized state and then again to a two cluster state, upon increasing the plasticity asymmetry parameter $\beta$. Thus the desynchronization state is always stable for the uniform distribution initial condition. Thus the two cluster-desynchronization-two cluster transition can also be termed as multi-stable $\rightarrow$ desynchronization $\rightarrow$ multi stable state.

We have also analytically investigated the linear stability of the desynchronized and the two cluster states in the limit $\eta\rightarrow0$ and have found the occurrence of multi-stable $\rightarrow$ desynchronization $\rightarrow$ multi-stable state transition. Our analytical results are in good agreement with our numerical observations.

We strongly believe that the model discussed here and the results therein will help fill the gap in the understanding of the dynamical aspects of adaptively coupled systems which are found to be most common in real world complex networks. This understanding will hopefully be helpful in elucidating the mechanism underlying the self-organizational nature of real world complex systems.

\section*{Acknowledgments}
The work is supported by the Department of Science and Technology (DST)--Ramanna program (ML), DST--IRHPA research project (ML, VKC and BS). ML is also supported by a DAE Raja Ramanna Fellowship. His work has also been supported by the Alexander von Humboldt Foundation, Germany under their Renewed Visit program to visit Potsdam where the work was completed. JHS is supported by a DST--FAST TRACK Young Scientist research project. JK acknowledges the support from LINC(EU ITN).

\end{document}